\documentclass[%
 twocolumn,
superscriptaddress,
 amsmath,amssymb,
 aps,
]{revtex4}

\usepackage{graphicx}
\usepackage{dcolumn}
\usepackage{bm}
\usepackage{xcolor}
\newcolumntype{.}{D{.}{.}{-1}}
\newcolumntype{d}[1]{D{.}{.}{#1}}
\newcommand*{\wn}{cm$^{-1}$}
\newcommand*{\hsm}{H$_{2}$S}

\begin{document}

\title{Two-photon Doppler-free ultraviolet laser spectroscopy on sulphur atoms}

\author{K.-F. Lai}
 \affiliation{Department of Physics and Astronomy, LaserLaB, Vrije Universiteit \\
 De Boelelaan 1081, 1081 HV Amsterdam, The Netherlands}
 \author{E. J. Salumbides}%
  \affiliation{Department of Physics and Astronomy, LaserLaB, Vrije Universiteit \\
 De Boelelaan 1081, 1081 HV Amsterdam, The Netherlands}
\author{W. Ubachs}%
 \affiliation{Department of Physics and Astronomy, LaserLaB, Vrije Universiteit \\
 De Boelelaan 1081, 1081 HV Amsterdam, The Netherlands}

\date{\today}

\begin{abstract}

The $3p^{4}$  $^{3}$P$_{J}$ - $3p^{3}4p$ $^{3}$P$_{J}$ transition in the sulphur atom is investigated in a precision two-photon excitation scheme under Doppler-free and collision-free circumstances yielding an absolute accuracy of 0.0009 \wn, using a narrowband pulsed laser. This verifies and improves the level separations between amply studied odd parity levels with even parity levels in S I. An improved value for the $^{3}$P$_{2}$ - $^{3}$P$_{1}$ ground state fine structure splitting is determined at $396.0564$ (7) \wn. A $^{34}$S - $^{32}$S atomic isotope shift was measured from combining time-of-flight mass spectrometry with laser spectroscopy.

\end{abstract}

\maketitle

\section{Introduction}

The odd parity level energies for the neutral sulfur atom have been extensively studied through VUV absorption spectroscopy from the ground electronic configuration \cite{kaufman1982,sarma1984,Joshi1987,zhou2008}. The connection with even parity excited states is studied through visible and infrared spectroscopy involving transition between excited states~\cite{frerichs1933,meissner1933,jakobsson1967,baclawski2011}. In addition, direct measurements of transitions between even parity states are studied through 2+1 resonance enhanced multiphoton ionization (REMPI) spectroscopy  \cite{Venkitachalam1991,woutersen1997}.

The level energies of the $^{1}$D$_{2}$ and $^{1}$S$_{0}$ states of the $3p^{4}$ ground electronic configuration, were investigated via electric dipole-forbidden transitions, first measured by McConkey et al. ~\cite{mcconkey1968}, and revisited at higher accuracy by Eriksson~\cite{eriksson1978}. Based on the combination differences between the forbidden transitions, $^{3}$P$_{1}$ - $^{1}$S$_{0}$, $^{1}$D$_{2}$ - $^{1}$S$_{0}$  and $^{3}$P$_{2}$ - $^{1}$D$_{2}$ the level energies of the lowest five levels are determined at an uncertainty of 0.005 \wn~\cite{eriksson1978}. 
Later, Brown et al. measured the fine structure transition $^{3}$P$_{1}$ - $^{3}$P$_{0}$ using laser magnetic resonance yielding an accuracy better than 10$^{-4}$ \wn~\cite{brown1994}.
The resulting level structure of the sulphur atom including a comprehensive compilation of lines and level energies is now well documented~\cite{martin1990,kaufman1993,morton2003}.

In the present study, high-resolution spectra of $3p^{4}$ $^{3}$P$_{J}$ - $3p^{3}4p$ $^{3}$P$_{J}$ transitions of $^{32}$S are measured by using 2+1 REMPI employing a narrowband pulsed laser amplifier in a scheme with counter-propagating laser beams, thus allowing for Doppler-free spectroscopy at high resolution and high accuracy.
The study is aimed at accurately bridging the large energy gap between the ground state and the manifold of excited states, which can be probed at high accuracy via infrared and visible spectroscopies. Via this means the measurement of a few transitions will allow for improving the accuracy of the entire level structure of the S atom. Moreover it will be shown that isotope shifts can be resolved in such Doppler-free precision experiment.

\section{Experiment}

\begin{figure}[b]
\begin{center}
\includegraphics[width=\linewidth]{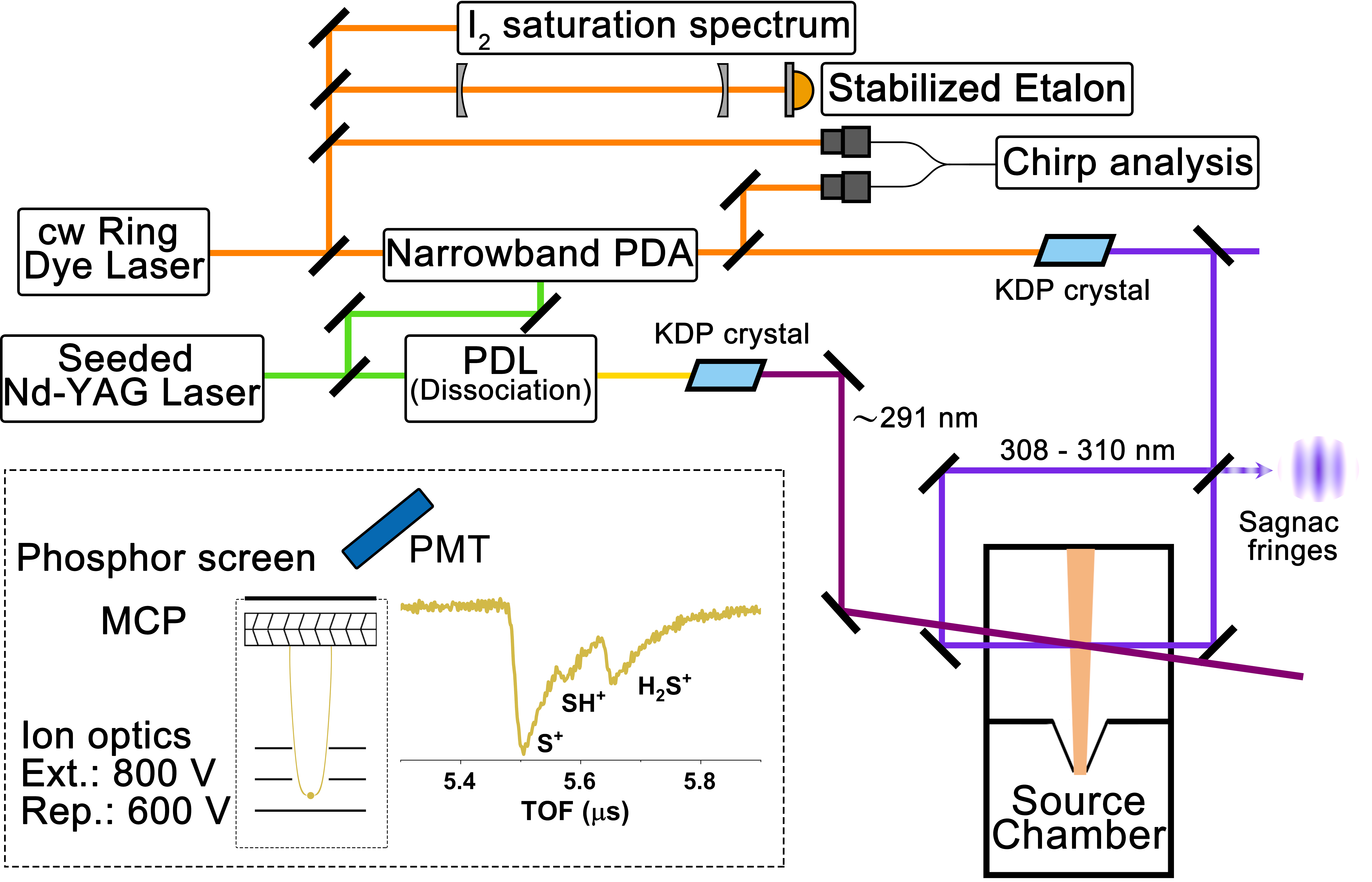}
\caption{\label{setup}
Schematic layout of the experimental setup. For details see text.}
\end{center}
\end{figure}

The experimental setup, schematically shown in Fig. \ref{setup}, is similar to that used for the production and detection of vibrationally excited states in molecular hydrogen, also obtained from photolysis of \hsm~\cite{niu2015,trivikram2016,trivikram2019}. Two ultraviolet (UV) pulsed laser systems are used to produce sulphur atoms and perform the precise two-photon spectroscopy. Sulphur atoms in the $3p^{4}$ $^{3}$P$_{J}$ ground state triplet are formed by UV-photodissociation of \hsm\ molecules, a well studied photolysis process~\cite{steadman1988,steadman1989,cook2001,zhou2020}. The first UV-laser pulse, inducing the dissociation, is obtained from a frequency-doubled pulsed dye laser (PDL) pumped by an injection-seeded pulsed Nd-YAG laser. Pulse energies of up to 4.5 mJ pulse are used for the photolysis. The wavelength of the dissociation laser is chosen at 291 nm following the original work of Steadman and Baer~\cite{steadman1988,steadman1989}.

The two-photon transition is measured by a traveling-wave pulsed-dye-amplifier (PDA) system amplifying the output of a narrowband cw-ring dye laser. The amplification is realized in three consecutive dye cells, pumped with the same Nd-YAG pump laser also used to pump the PDL~\cite{Eikema1997}. The output of the PDA at 616 - 621 nm is frequency-doubled in a KDP crystal to provide UV-pulses in the range 308 - 311 nm with $\sim 4$ ns pulse width.
The frequency of the cw-seed light is calibrated against the standard of I$_2$ saturated hyperfine lines combined with the transmission markers of a stabilized Fabry-Perot interferometer~\cite{xu2000}. The chirp effect on the pulses, giving rise to an effective frequency offset between the pulsed output of the PDA and the cw-light is assessed via optical heterodyne measurements and analyzed via known techniques~\cite{melikechi1994,Eikema1997}.  The narrowband UV beam is then split and configured in a counter-propagating beam setup to induce the Doppler-free two-photon transitions. The angle mismatch of the counter-propagating beams is reduced based on Sagnac interference fringes~\cite{Hannemann2007}.

\begin{table}[b]
\renewcommand{\arraystretch}{1.3}
\caption{\label{tab:transition}
Measured frequencies for the two-photon transitions $3p^{4}$ $^{3}$P$_{J}$ - $3p^{3}4p$ $^{3}$P$_{J}$ of $^{32}$S, with uncertainties indicated in parentheses.}
\begin{tabular}{ccl}
Initial state	& Excited state	& \multicolumn{1}{c}{Obs. (\wn )} \\
\hline
& $^{3}$P$_{1}$  &64888.9317\,(9) \\
$^{3}$P$_{2}$   & $^{3}$P$_{0}$  &64891.3536\,(9) \\
& $^{3}$P$_{2}$  &64892.5494\,(9) \\
\hline
$^{3}$P$_{1}$   & $^{3}$P$_{1}$  &64492.8751\,(9) \\
& $^{3}$P$_{2}$  &64496.4917\,(23) \\
\hline
$^{3}$P$_{0}$   & $^{3}$P$_{0}$  &64317.7561\,(9) \\
& $^{3}$P$_{2}$  &64318.9561\,(9) \\
\hline
\end{tabular}
\end{table}

The UV beams are focused into a spot of size few tens of $\mu$m spatially overlapping a pulsed \hsm\ beam, in a low-density region of a skimmed and collimated pulsed effusive gas expansion. To avoid ac-Stark disturbances from the dissociation laser, the PDA spectroscopy laser is optically delayed by 10 ns with respect to the photolysis laser, such that there is no temporal overlap. Sulphur atom signal is generated via 2+1 REMPI, whereby ions are extracted through a mass-resolving time-of-flight (TOF) tube, detecting S$^+$ ions. Ion optics are triggered at a delay of $\sim 50$ ns from the spectroscopy laser, so that the laser-excitation takes place in zero DC field. The ion signal is amplified by a microchannel plate (MCP) with phosphor imaging screen with detection on a photomultiplier tube. Mass-selected spectra are recorded with a box-car integrator probing only a narrow channel of the TOF-trace. The large amounts of SH$^+$ and H$_2$S$^+$ signal in nearby mass channels, as well as S$^+$ background signals from various dissociation/ionization channels are limiting factors on the signal-to-noise-ratio of the S-atom spectra. In case of spectral recording of measuring spectra of $^{34}$S this is even more detrimental.

\begin{figure}[t]
\begin{center}
\includegraphics[width=\linewidth]{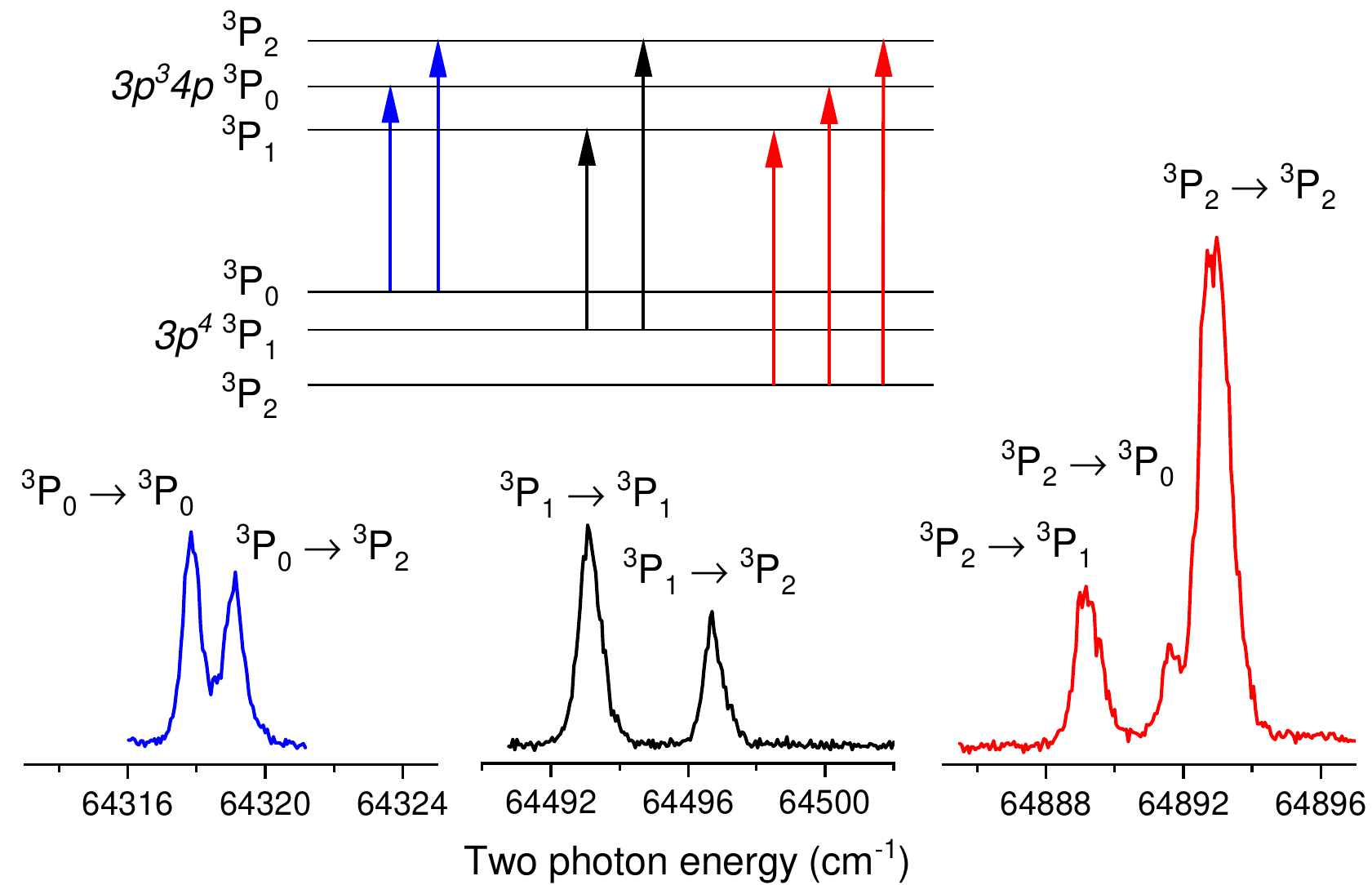}
\caption{\label{overview}
Recorded two-photon spectra of all seven components in the $3p^{4}$  $^{3}$P$_{J}$ -  $3p^{3}4p$ $^{3}$P$_{J}$ multiplet under Doppler-broadened conditions. The inset shows a level diagram connecting the levels probed.
}
\end{center}
\end{figure}

\section{Results and Interpretation}

All of the seven two-photon allowed transitions between $3p^{4}$ $^{3}$P$_{J}$ and $3p^{3}4p$ $^{3}$P$_{J}$ were measured in the wavelength interval 308-311 nm. Figure~\ref{overview} displays recordings of all observed lines under Doppler-broadened conditions. Note that the combination $J= 0 \leftrightarrow 1$ is forbidden by two-photon selection rules~\cite{Dixit1988}.

\begin{figure}[t]
\begin{center}
\includegraphics[width=\linewidth]{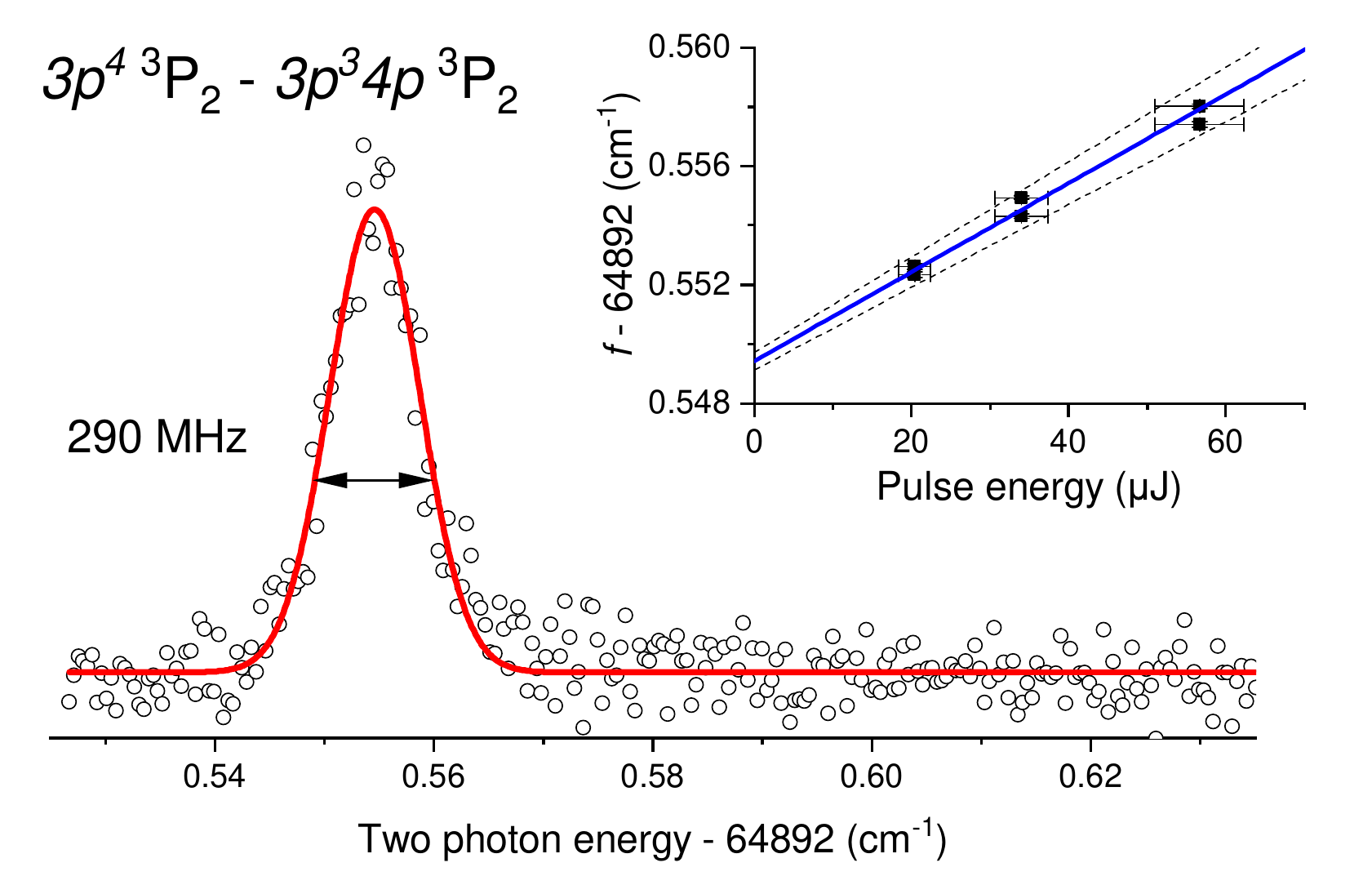}
\caption{\label{3p2_3p2}
Spectrum of  $3p^{4}$  $^{3}$P$_{2}$ -  $3p^{3}4p$ $^{3}$P$_{2}$ two-photon transition of S I recorded at lowest power and fitted with Voigt function. The inset shows the power dependence of the transition frequency denoted with $f$.
}
\end{center}
\end{figure}

\begin{figure}[t]
\begin{center}
\includegraphics[width=\linewidth]{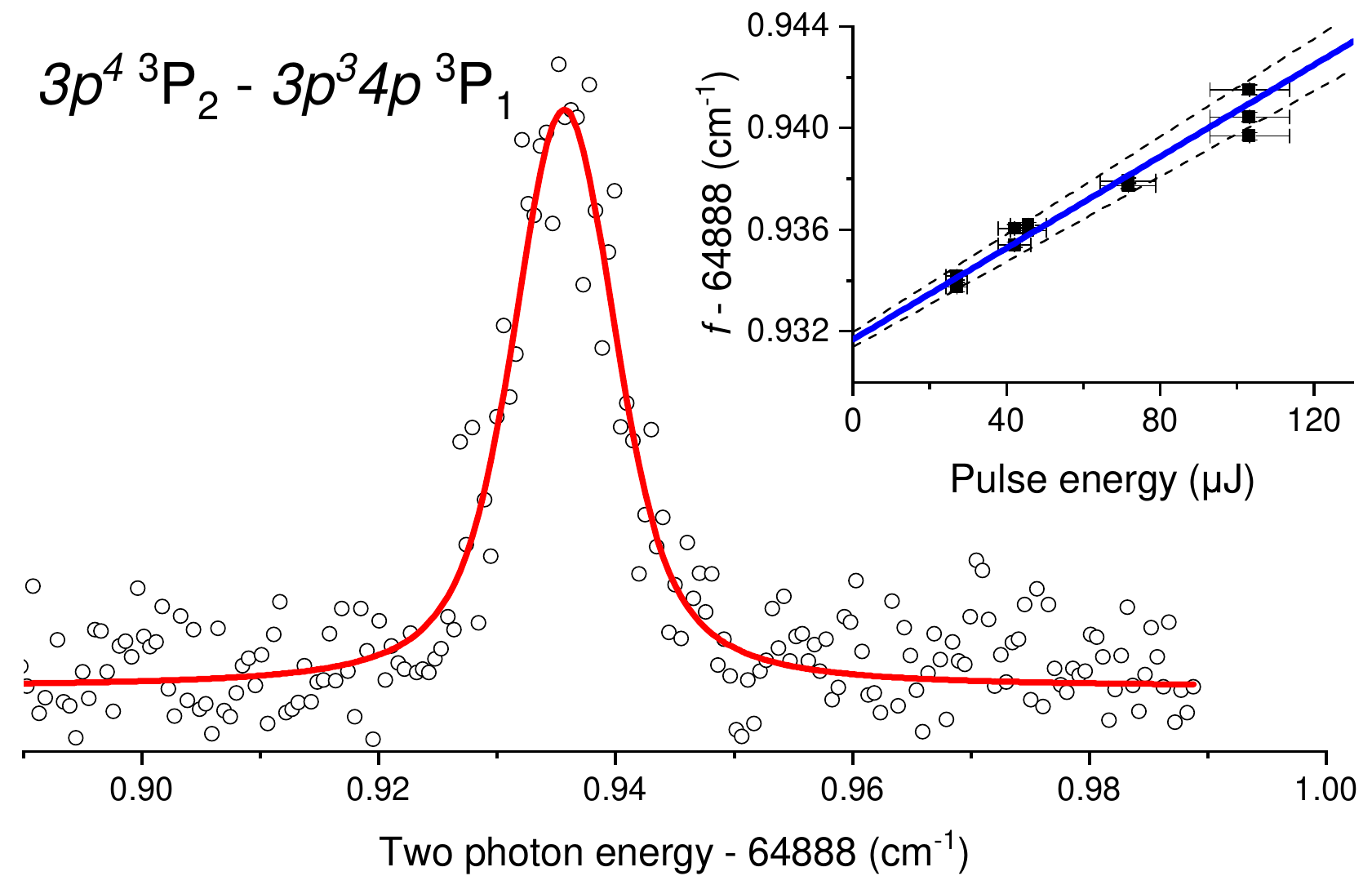}
\caption{\label{3p1_3p2}
Spectrum of $3p^{4}$ $^{3}$P$_{2}$ -  $3p^{3}4p$ $^{3}$P$_{1}$ two-photon transition of S I; same details as Fig.~\ref{3p2_3p2}.}
\end{center}
\end{figure}

The spectra for the $3p^{4}$ $^{3}$P$_{2}$ - $3p^{3}4p$ $^{3}$P$_{2}$ and $3p^{4}$ $^{3}$P$_{2}$ - $3p^{3}4p$ $^{3}$P$_{1}$ lines, recorded under Doppler-free conditions, are shown in greater detail in Figs.~\ref{3p2_3p2} and \ref{3p1_3p2}.  The width of the spectral lines, measured at the lowest power, is about 290 MHz (FWHM), only slightly larger than expected by assuming exact Fourier-transform limited laser pulses of Gaussian spectral profile. The power dependence (or ac-Stark effect) for the transition frequencies is studied by varying the PDA pulse energy as shown in the inset of the figures.
Table~\ref{tab:transition} lists the transition frequencies, upon extrapolation to zero field, as measured for the seven transitions with the boxcar gate set to $^{32}$S.

\begin{table}
\renewcommand{\arraystretch}{1.2}
\caption{\label{tab:error}
Error budget for the two-photon frequencies for the S atom measured in the present study, except for the $3p^{4}$ $^{3}$P$_{1}$ - $3p^{3}4p$ $^{3}$P$_{2}$ line, where the uncertainty is larger.
}

\begin{tabular}{lc}
Contribution & Uncertainty ($\times 10^{-4}$ \wn ) \\
\hline
Line profile (fitting) & 2\\
Statistics & 3\\
AC-Stark extrapolation & 5 \\
Frequency calibration & 3\\
Cw-pulse offset (chirp)  & 6\\
Residual Doppler & $< 1$\\
DC-Stark effect & $< 1$\\
\hline
Total & 9\\
\end{tabular}
\end{table}

The sources of uncertainty are summarized in an error budget presented in Table \ref{tab:error}.
A statistical analysis of the determination of the line centres gives an uncertainty of $3  \times \ 10^{-4}$ \wn, including averaging over multiple recordings. The ac-Stark effect is the dominant systematic effect in the present study. It causes a shift of line centres, accompanied by broadening, and due to the spatial variation of laser intensity over the laser focus, also results in an asymmetry of the line profile~\cite{trivikram2016}.  The asymmetry was addressed by fitting skewed Voigt profile fitting. Analysis of the line shape results in an additional contribution to the uncertainty of $2 \times \ 10^{-4}$ \wn. The ac-Stark shift is further analyzed by performing measurements over a range of pulse energies of 20 - 100 $\mu$J with extrapolation of the centre frequency to zero energy. This adds a contribution to the error budget of $5 \times 10^{-4}$ \wn.
Further contributions are associated with the frequency chirp in the PDA-system and the absolute frequency calibration against I$_2$-hyperfine components, which were analyzed by established techniques~\cite{ubachs1997,Eikema1997} and result in a contribution of $6 \times 10^{-4}$ \wn\ for the frequency uncertainty. The absolute frequency calibration against I$_2$ hyperfine components involves uncertainty in the reference frequencies~\cite{xu2000} and measurement of the FSR, amounting to $3 \times 10^{-4}$ \wn.
For the latter two contributions multiplication by four, for the frequency doubling and the two-photon process, is included.
The experiment is essentially Doppler-free, although small shifts of the frequency centre may be associated with a non-isotropic velocity distribution of the S-atoms, similar to the case of H$_2$ investigated~\cite{cheng2018}.
For this reason the counter-propagating laser beams were aligned in a Sagnac interferometer~\cite{Hannemann2007} limiting this effect to below $1 \times 10^{-4}$ \wn. Excitation was performed in zero field, hence the DC-Stark effect is negligible on the scale of the present accuracy.
Taking the contributions in quadrature leads to a total uncertainty of 0.0009 \wn\ for the frequencies of the two-photon resonances for all observed transitions except for one. The uncertainty of $3p^{4}$ $^{3}$P$_{1}$ - $3p^{3}4p$ $^{3}$P$_{2}$ is estimated at 0.0023 \wn\ with larger uncertainty from statistics, a long measurement trace to be covered for reaching an I$_2$ resonance, and problems encountered in ac-Stark extrapolation.

\begin{figure}[b]
\begin{center}
\includegraphics[width=0.9\linewidth]{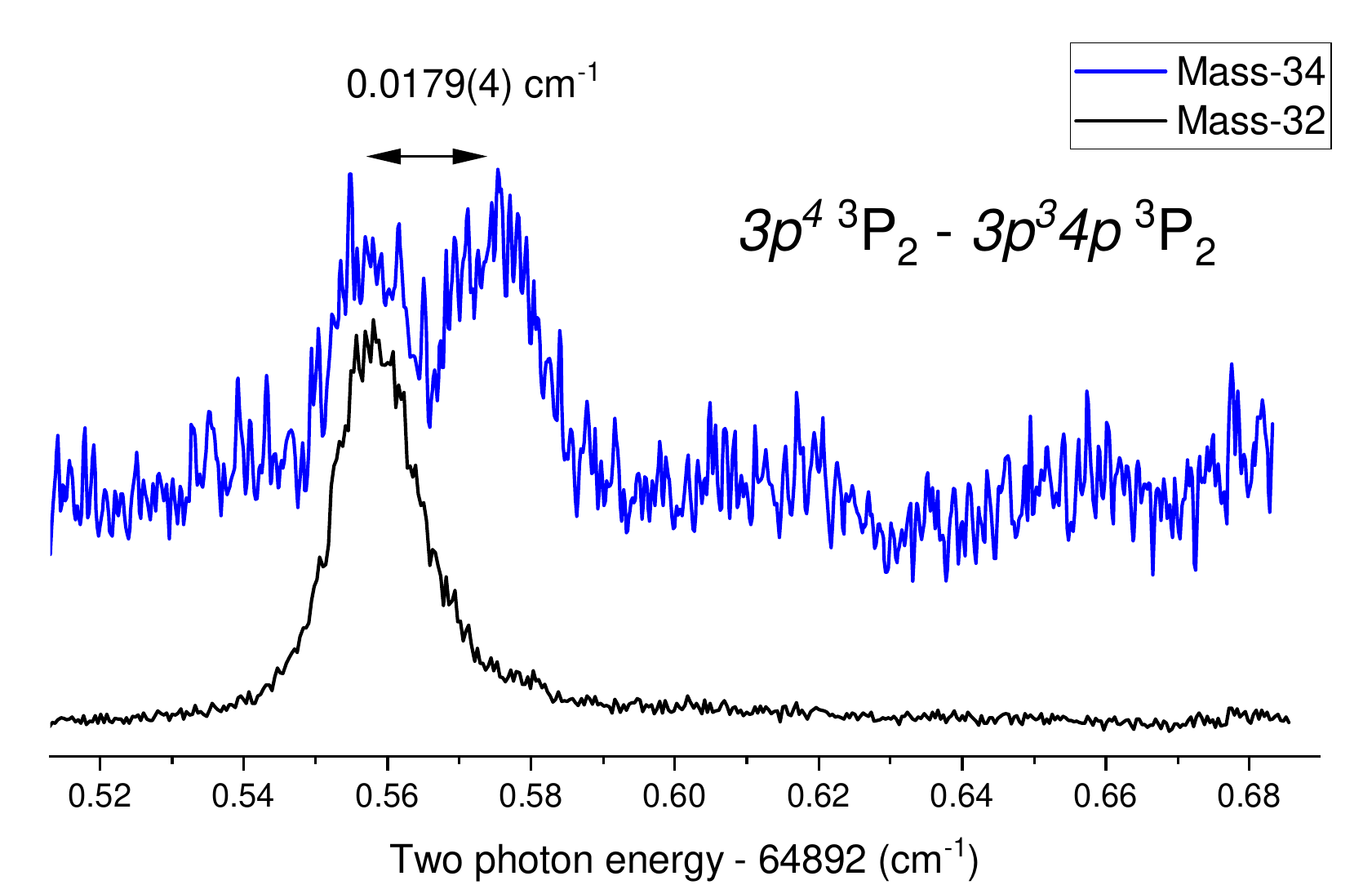}
\caption{\label{s34}
Isotope shift in the $3p^{4}\,{}^{3}\mathrm{P}_{2}\rightarrow 3p^{3}4p\,{}^{3}\mathrm{P}_{2}$ transition of the $^{34}$S isotope (blue curve) recorded by gating on the mass-$34$ channel, which coincides with the H$_2$S$^{+}$ signal. Resonance signal from the main isotope $^{32}$S is still observable in the mass-$34$ channel, overlapping the spectrum observed in mass-$32$ channel (black curve).
}
\end{center}
\end{figure}

In a single case, for the  strongest line $^{3}$P$_{2}$ - $^{3}$P$_{2}$, a study was made of the resonance line in $^{34}$S from the 5\% naturally abundant isotope in the sample. The spectrum, shown in Fig.~\ref{s34}, and recorded with a boxcar gate  probing mass-34, displays the much lower signal-to-noise ratio, caused by spurious signal on mass-34 of H$_2$S$^+$ ions. When only considering the statistical line fitting and relative calibration errors the  isotope shift on the resonances amounts to 0.0179 (4) \wn.
The spectrum of Fig.~\ref{s34} shows that the spectral contributions of  $^{32}$S and $^{34}$S are well separated, thus verifying that the listed entries for the transition frequencies in Table \ref{tab:transition} pertain to the main $^{32}$S isotope, and do not correspond to a mixture of isotopes.

The isotope shift for a transition can be separated into normal mass shift, specific mass shift and field shift contributions~\cite{Carette2010}.
In view of the very small wave function amplitude of the $p$ outer electron within the nuclear charge radius, the field shift (or finite size) contribution is negligible for the differential isotope effect.
The Bohr shift or normal mass shift ($\Delta E_{\rm NMS} = E_{\rm NMS}(^{34}\mathrm{S})-E_{\rm NMS}(^{32}\mathrm{S})$) is calculated to be 0.0650~\wn.
From the experimental isotope shift obtained here, a specific mass shift of $\Delta E_{\rm SMS}$=-0.0471(4)~\wn\ is extracted for the $3p^{4}\,\,^{3}\mathrm{P}_{2}\rightarrow 3p^{3}4p\,\,^{3}\mathrm{P}_{2}$ transition. The $\Delta E_{\rm SMS}$ experimental value will be useful in validating \emph{ab initio} calculations of electron correlations.

\section{Discussion: Level energies}

\begin{table}
\renewcommand{\arraystretch}{1.3}
\caption{\label{tab:level}
Least-square fitting for level energy of $3p^{3}4p$ $^{3}$P$_{J}$ and $3p^{4}$ $^{3}$P$_{J}$ of $^{32}$S, with uncertainties relative to ground $3p^{4}$ $^{3}$P$_{2}$ state indicated in parentheses. A comparison is given with values given in Ref. \cite{martin1990}; for the quoted uncertainties see text.
All values are given in \wn.
}
\begin{tabular}{rccc}
Level	& \multicolumn{1}{c}{This work}	& \multicolumn{1}{c}{Ref. \cite{martin1990}} & \multicolumn{1}{c}{Difference} \\
\hline
$3p^{4}$ $^{3}$P$_{2}$  &\multicolumn{1}{c}{0} & \multicolumn{1}{c}{0} & - \\
$3p^{4}$ $^{3}$P$_{1}$ & 396.0570\,(11) &396.055\,(5) & 0.0020\\
$3p^{4}$ $^{3}$P$_{0}$ & 573.5953\,(9) &573.640\,(16) & -0.0447\\
\hline
$3p^{3}4p$ $^{3}$P$_{1}$ &64888.9319\,(9)  &64888.964\,(25)  & -0.0321\\
$3p^{3}4p$ $^{3}$P$_{0}$ & 64891.3525\,(8)  &64891.386\,(25) & -0.0335\\
$3p^{3}4p$ $^{3}$P$_{2}$ & 64892.5503\,(7) &64892.582\,(25) & -0.0317\\
\hline
\hline
& \multicolumn{1}{c}{This work} & \multicolumn{1}{c}{Ref. \cite{martin1990}} & \multicolumn{1}{c}{Ref. \cite{brown1994}} \\
\hline
$^{3}$P$_{0}$ - $^{3}$P$_{1}$ & \multicolumn{1}{c}{177.5383\,(14)} & \multicolumn{1}{c}{177.585\,(17)} & \multicolumn{1}{c}{177.539253\,(93)}\\
\hline
\end{tabular}
\end{table}

\begin{table}[t]
\renewcommand{\arraystretch}{1.3}
\caption{\label{tab:full level}
Least-square fitting for level energy of ground electron configuration $3p^{4}$ of $^{32}$S, with uncertainties indicated within parentheses. The transition from Ref.~\cite{eriksson1978,brown1994} and this work are included in fitting, and a comparison is made with the data compilation of Ref.~\cite{martin1990}. All values are given in \wn.
}
\begin{tabular}{r..}
\multicolumn{1}{c}{Level}	& \multicolumn{1}{c}{This work} & \multicolumn{1}{c}{Ref. \cite{martin1990}}\\
\hline
 $^{3}$P$_{2}$  &\multicolumn{1}{c}{0} & \multicolumn{1}{c}{0}  \\
 $^{3}$P$_{1}$ & 396.0564\,(7) &396.055\,(5)\\
 $^{3}$P$_{0}$ & 573.5956\,(7) &573.640\,(16)\\
 $^{1}$D$_{2}$ & 9238.6097\,(23)  &9238.609\,(5) \\
 $^{1}$S$_{0}$ & 22179.9548\,(22)  &22179.954\,(5) \\
 \hline
\end{tabular}
\end{table}

The precise determination of transition frequencies can be cast into a least-squares analysis to determine level energies in both $^{3}$P states in the $3p^{4}$ and $3p^{3}4p$ configurations using the LOPT program \cite{kramida2011}. First the internal consistency of the measurements can be tested by including only the presently obtained data set, as is done in Table~\ref{tab:level}. The LOPT analysis provides a consistent set of level energies at an accuracy in most cases below $10^{-4}$ \wn. More importantly the deduced ground state splitting between $^{3}$P$_{1}$ and $^{3}$P$_{0}$ levels is in full agreement (within $1\sigma$) with the very accurate LMR-measurement~\cite{brown1994}, the most precise level splitting determined in the sulphur atom. This agreement provides proof that the uncertainties of the present study are not underestimated.

In Table~\ref{tab:level} also a comparison is made with the results from VUV spectroscopy, which are at the basis of the comprehensive published line and level lists~\cite{kaufman1982,martin1990,kaufman1993}. In these compilations a single spectral line in the VUV is included (a transition to the $4s$ $^3$S$_1$ level) for which an uncertainty as low as $0.025$ \wn\ is stated~\cite{kaufman1982}. We adopt this value as the general uncertainty for the level energies for the excited states, although the uncertainty for the overall level structure might be somewhat larger.
The increased precision on the  $^{3}$P ground state level energies in the compilations, so better than the quoted $0.025$ \wn, in fact derive from the rather accurate measurement of the forbidden transitions by Eriksson~\cite{eriksson1978}.
The $^{3}$P$_{0}$ level was not accessed in the measurement of Ref.~\cite{eriksson1978}, hence its uncertainty relies on VUV-data.
Viewed in this context the deviations between present results and the VUV-compilation~\cite{kaufman1982}, as listed in Table~\ref{tab:level} both for ground state splittings and $3p^34p$ excitation energies are close to the expected uncertainties. This includes the consistent shift -0.032 \wn\ for all three levels in the $3p^{3}4p$ $^{3}$P$_{J}$ excited triplet.
This finding is indicative for an overall systematic shift of all excited level in the data compilation by $0.03$ \wn.



Finally a LOPT least-squares analysis can be performed to determine the level energies of the entire $3p^4$ ground electronic configuration including $^{1}$D$_{2}$ and $^{1}$S$_{0}$ levels, based on the present study in combination with the high precision measurements of magnetic dipole transitions in Ref.~\cite{eriksson1978,brown1994}. Table \ref{tab:full level} lists the fitted level energy values with individual uncertainties relative to ground state and comparison with the corresponding values listed in the S-atom data compilation~\cite{martin1990}. This results in an improved level structure for $3p^4$, in particular for the lowest fine structure splitting $^{3}$P$_{2}$ - $^{3}$P$_{1}$ which is determined at $396.0564$ (7) \wn, corresponding to a far-infrared wavelength of 25.24893 (4) $\mu$m.

\section{Conclusion}

In conclusion, seven transitions in the  of $3p^{4}$ $^{3}$P$_{J}$ -  $3p^{3}4p$ $^{3}$P$_{J}$ multiplet are measured by narrowband laser spectroscopy at an uncertainty of 0.0009 \wn. For the first time a $^{34}$S - $^{32}$S isotope shift has been measured in atomic sulphur, from which a value for the specific mass-shift was derived, a measure for electron correlations in the atom. The accurate transition frequencies improve the level energies of the $3p^{4}$ $^3$P ground electronic configuration by factor of two. The $3p^{3}4p$ $^{3}$P$_{J}$ excited state level energies are determined at an absolute accuracy of less than 0.001 \wn. The present study provides an indication of an overall systematic shift for the excited level energies as listed in spectroscopic data compilations for the sulphur atom~\cite{kaufman1982,martin1990,kaufman1993}. The precise measurement of even parity excited states may help optimizing the level energies of odd parity levels by future improved measurements between excited states in the infrared and visible regions, therewith using the  $3p^{3}4p$ $^{3}$P$_{J}$ levels as anchor levels, in a similar fashion as applied to H$_2$~\cite{bailly2010}.

\section*{Acknowledgement}

WU acknowledges the European Research Council for an ERC Advanced grant (No: 670168).

%

\end{document}